\documentclass[superscriptaddress, reprint, amsmath, amssymb, aps, pra, floatfix]{revtex4-2}
\usepackage[colorlinks=true]{hyperref}
\usepackage{graphicx}
\usepackage{dcolumn}
\usepackage{bm}
\usepackage{orcidlink}
\usepackage{physics}
\usepackage{color}
\usepackage{amsthm}
\usepackage{multirow}
\usepackage{comment}

\usepackage{xcolor}
\definecolor{pink}{rgb}{1,0,0.9}

\newcommand{\change}[1]{{#1}}
\newcommand{\changeb}[1]{{#1}}
\usepackage[normalem]{ulem}

\begin{document}

\title{Enhancement of superconducting pairing via quantum \change{Lyapunov} control}

\author{Oleksandr Povitchan%
\orcidlink{0009-0007-6199-9857}
}\email{opovitchan@student.ethz.ch}
\affiliation{Department of Physics, ETH Z\"urich, Otto-Stern-Weg 1, 8093 Z\"urich, Switzerland}
\affiliation{Education and Research Institute ``School of Physics and Technology'', Karazin Kharkiv National University, Svobody Square 4, 61022 Kharkiv, Ukraine}

\author{Denys I. Bondar%
\orcidlink{0000-0002-3626-4804}
}\email{dbondar@tulane.edu}
\affiliation{Department of Physics and Engineering Physics, Tulane University, 6823 St. Charles Avenue, New Orleans, Luisiana 70118, USA}

\author{Andrii G. Sotnikov%
\orcidlink{0000-0002-3632-4790}
}\email{a\_sotnikov@kipt.kharkov.ua}
\affiliation{Education and Research Institute ``School of Physics and Technology'', Karazin Kharkiv National University, Svobody Square 4, 61022 Kharkiv, Ukraine}
\affiliation{Akhiezer Institute for Theoretical Physics, NSC KIPT, Akademichna 1, 61108 Kharkiv, Ukraine}

\begin{abstract}
We demonstrate that \change{quantum Lyapunov} control provides an effective strategy for enhancing superconducting correlations in the Fermi-Hubbard model without requiring careful parameter tuning. While photoinduced superconductivity is sensitive to the frequency and amplitude of a monochromatic laser pulse, our approach employs a simple feedback-based protocol that prevents the decrease of superconducting correlations once they begin to form. This method enables robust enhancement of  pairing across a broad range of initial pumping conditions, eliminating the need for intricate frequency and amplitude optimization. We also show that an alternative implementation, asymptotic quantum control, achieves comparable results. Furthermore, our approach can be adapted to suppress previously induced superconducting correlations, providing bidirectional control over quantum pairing states. These findings suggest practical pathways for manipulating quantum correlations in strongly interacting systems with minimal experimental complexity.
\end{abstract}

\maketitle

\section{Introduction}

Superconducting long-range order~\cite{Yang1962APS}, present in many eigenstates of the Hubbard Hamiltonian, has been known for more than three decades~\cite{Yang1989PRL}. Recent theoretical studies~\cite{Aoki2014,Sentef2016,Kaneko2019PRL,Kaneko2020PRL}, inspired by the rich nonequilibrium dynamics observed in condensed-matter systems~\cite{Fausti2011, Mitrano2016, Cantaluppi2018}, demonstrated that an external driving field can induce superconducting-like behavior even in the Mott-insulating regimes of half-filled Fermi-Hubbard chains. This driving gives rise to the so-called $\eta$ pairing of interacting fermions in one-dimensional chains~\cite{Yang1989PRL}, analogous to the formation of conventional $s$-wave-type ordered states in superconductors. Such excitation generates nonvanishing charge stiffness and long-range pairing correlations, as detailed in Refs.~\cite{Kaneko2019PRL,Kaneko2020PRL}. In these works, the researchers utilized monochromatic laser pulses with a Gaussian envelope and identified specific frequencies and amplitudes that enhance the system's superconducting properties. However, determining these optimal pulse parameters remains challenging. The interacting Fermi-Hubbard chain is nearly transparent at most frequencies, and even at resonant frequencies, an improperly chosen pulse amplitude can cause rapid decay of correlations before the field is completely turned off.

In this paper, we show that $\eta$ pairing can be efficiently manipulated by \change{quantum Lyapunov} control~\cite{kosloff_excitation_1992, sugawara_control_1994, sugawara_control_1995, ohtsuki_application_1998, tannor_laser_1999, sugawara_general_2003, mirrahimi_reference_2005} to enhance and suppress superconductive correlations at will. This study shows the efficiency in dynamic manipulation of the driving field to enhance superconducting pairing.
The key finding is that shaping \change{the pulse envelope with automatic dynamical adjustment of the carrier frequency in the low-field regime} yields much better enhancement than using a Gaussian envelope with a fixed carrier frequency. The quantum Lyapunov control is a time-local method, meaning that the control-field value is calculated at every time step. This computationally efficient procedure requires only a single-pass solution of the Schr\"odinger equation. It also contrasts with the optimal quantum control~\cite{Koch_22, morzhin_krotov_2019, glaser2015training, petersen_quantum_2010, brif_control_2010, dalessandro_introduction_2007} which, although it is able to find better results, requires solving the Schr\"odinger equation tens or hundreds of times with different iterations of the sought control field. This advantage of \change{quantum Lyapunov} control recently opened new venues in quantum computing~\cite{magann_lyapunov-control-inspired_2022, larsen_feedback-based_2024}.

The developed methodology can be applied to ultracold atomic quantum simulators of Hubbard models to explore nonequilibrium dynamics and pairing mechanisms of interacting fermions under highly controlled conditions~\cite{Bloch2012}. With further extensions to nonzero temperature, dissipation, and higher-dimensional lattice systems, it has the potential to describe 
\changeb{new generations of experiments with optical pulse engineering oriented on the enhancement of superconducting correlations in solid-state materials both below and above the equilibrium critical temperature of superconducting pairing.}

\section{Photoinduced superconductivity in the Hubbard Model}

\subsection{System under study}

We consider a half-filled spin-balanced (the total numbers of spin-up and spin-down particles are equal, $N_{\uparrow}^{}=N_{\downarrow}=L/2$) one-dimensional Fermi-Hubbard chain of length~$L$ with the periodic boundary conditions described by the Hamiltonian
\begin{equation} \label{eq:hamiltonian}
    \hat{\mathcal{H}} = -t_h \sum_{i,\sigma}(e^{i\Phi(t)}\hat{c}_{i,\sigma}^{\dagger} \hat{c}_{i+1,\sigma}^{} + \text{H.c.}) + U \sum_{i} \hat{n}_{i,\uparrow}^{} \hat{n}_{i,\downarrow}^{},
\end{equation}
with the driving field $\Phi(t)$ introduced via the Peierls substitution. The creation and annihilation operators satisfy the standard fermionic anticommutation relation $\{\hat{c}_{i,\nu}^{\dagger},\hat{c}_{j,\mu}^{}\} = \delta_{i,j}^{}\delta_{\nu,\mu}^{}$, while the density of electrons in the spin state $\sigma=\{\uparrow,\downarrow\}$ on lattice site $i$ is determined by the operator $\hat{n}_{i,\sigma}^{} = \hat{c}_{i,\sigma}^{\dagger}\hat{c}_{i,\sigma}^{}$. The Hamiltonian~\eqref{eq:hamiltonian} and all other quantities below are expressed in atomic units (the reduced Planck constant $\hbar$, the elementary charge~$e$, the electron mass $m_e$, and the inverse Coulomb constant $4\pi\varepsilon_0$ are all set  to 1). We restrict ourselves to the Mott-insulating regime with $U/t_h = 20$, where $U$ is the onsite interaction strength, and the chain length $L = 8$. The hopping amplitude $t_h$ and the lattice constant $a$ are taken below as the scaling units (i.e., $t_h=1$ and $a=1$).

\subsection{Order parameter}

To examine whether the system possesses nontrivial superconducting correlations, we track the order parameter $\langle\hat{\eta}^2\rangle/L$, defined as the expectation value of the operator
\begin{equation} \label{eq:eta_sq}
    \hat{\eta}^2_{} = \frac{1}{2} ( \hat{\eta}^{+}_{} \hat{\eta}^{-}_{} + \hat{\eta}^{-}_{} \hat{\eta}^{+}_{}  ) + \hat{\eta}_{z}^{2}
\end{equation}
normalized by the system size $L$. 
The ladder-type operators $\hat{\eta}^{+}_{} = \sum_{i} (-1)^i \hat{\eta}^{+}_{i}$ (constructed in terms of the pair-creation operators $\hat{\eta}^{+}_{i}=\hat{c}_{i,\downarrow}^{\dagger} \hat{c}_{i,\uparrow}^{\dagger}$) and $\hat{\eta}^{-}_{} = (\hat{\eta}^{+}_{})^{\dagger}$, together with $\hat{\eta}^{}_{z} = \frac{1}{2} \sum_{i} ( \hat{n}_{i,\uparrow} + \hat{n}_{i,\downarrow} - 1  )$ satisfy the conventional commutation relations of the symmetry group SU(2). 

In the absence of external driving, $\hat{\eta}^2_{}$ and $\hat{\eta}^{}_{z}$ commute with the Hamiltonian~\eqref{eq:hamiltonian}. Let $\ket{\psi_m}$ denote the eigenstate of the  time-independent Hamiltonian~\eqref{eq:hamiltonian}, 
 \begin{align}\label{EqEigenstatesEigenenergies}
        \left. \hat{\mathcal{H}} \right|_{\Phi(t) = 0} \ket{\psi_m} = \varepsilon_m \ket{\psi_m},
        \quad 
        m = 1,\ldots, \binom{L}{L/2}^2.
 \end{align}
For each eigenstate $\ket{\psi_m}$, Fig.~\ref{fig:excited(FE)_eigens}(a) shows a black dot representing its corresponding energy $\varepsilon_m$ and the value of the order parameter $\langle\hat{\eta}^2\rangle_m = \bra{\psi_m} \hat{\eta}^2 \ket{\psi_m}$. \change{The colored diamonds in Fig.~\ref{fig:excited(FE)_eigens}(a) are discussed in the next section. The clustering of black dots around certain energy values visualizes the emergence of energy bands.

From Fig.~\ref{fig:excited(FE)_eigens}(a) one can also note that the ground state~$\ket{\psi_1}$ of the unperturbed system has a zero order parameter,  $\langle\hat{\eta}^2\rangle_1=0$.} This is a general observation, which is independent of the system size~$L$, and a consequence of the Mermin-Wagner theorem forbidding spontaneous breaking of the continuous symmetry of the many-body ground state in one-dimensional systems even at zero temperature~\cite{MerminWagner1966,Altland_Simons_2010}.

\subsection{Superconducting correlations induced by monochromatic pulses}\label{SecmonochromaticPulse}

Assume the system~\eqref{eq:hamiltonian} is initially in the ground state, $\ket{\psi_1}$, of the field-free Hamiltonian before being exposed to an external laser with electric field $E(t) = -\frac{1}{a}\frac{d \Phi(t)}{dt}$, 
\begin{equation} \label{eq:field}
    \Phi(t) = \Phi_0 \sin (\omega_p \tau) \sin^{2}  \left( \frac{\omega_p \tau}{2 N_p}\right) \phi(\tau),
\end{equation}
where the multiplier $\phi(\tau) = \theta(\tau) - \theta(\tau - T_pN_p)$, constructed by the two Heaviside functions, ensures exactly $N_p$ oscillations with period $T_p=2\pi/\omega_p$ under the single arc of the envelope $\sin^{2}  \left( \frac{\omega_p \tau}{2 N_p}\right)$. The evolution of the many-body wave function $|\Psi(t)\rangle$ is obtained by solving the time-dependent Schr\"odinger equation $\frac{\dd}{\dd t}|\Psi(t)\rangle = -i \hat{\mathcal{H}}(t)|\Psi(t)\rangle$ numerically with the time step $\delta t=0.02T_p$~\cite{QS2017,QS2019}.

\begin{figure}[tb]
    \includegraphics[width= \linewidth]{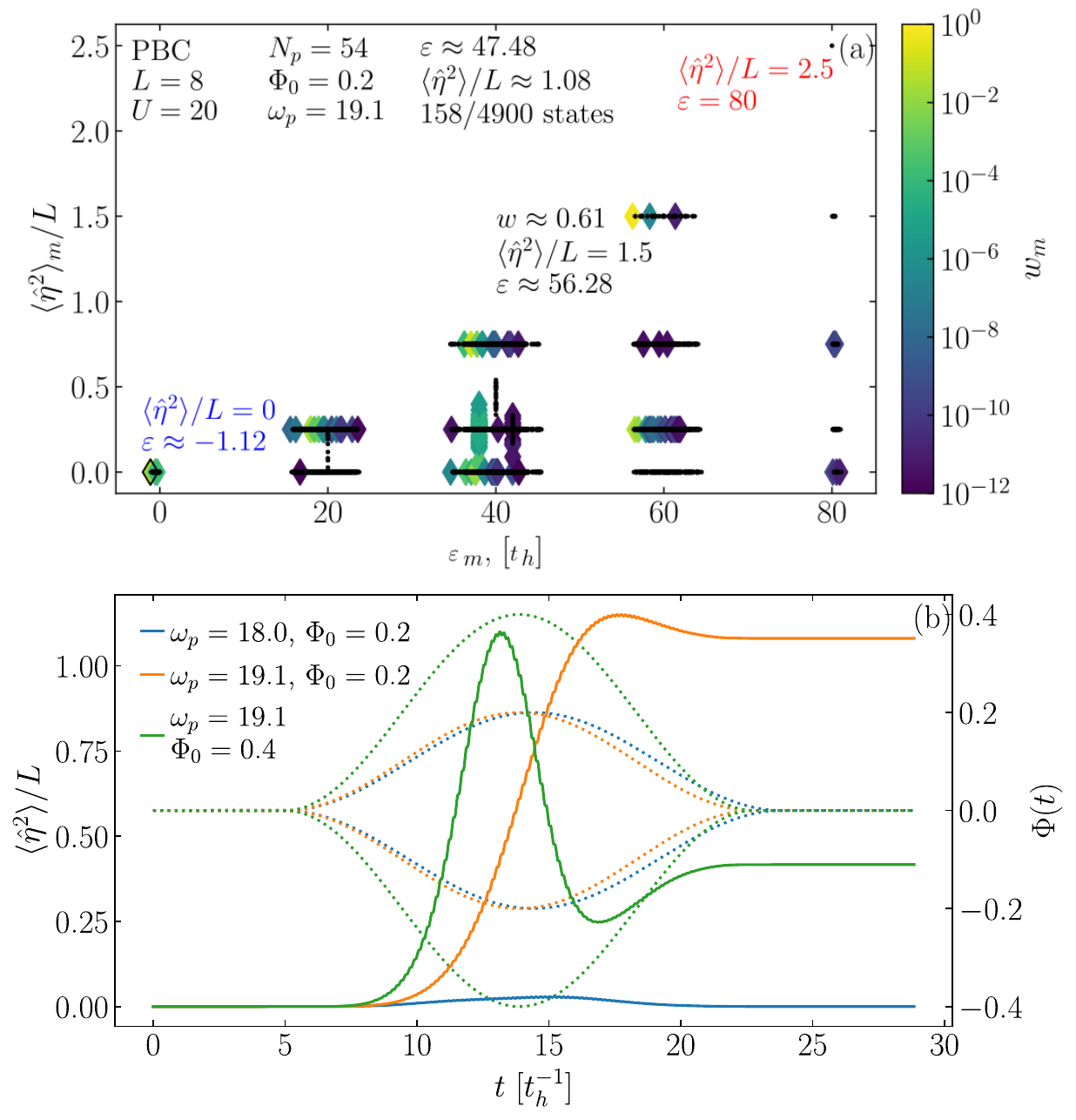}
    \caption{
    \label{fig:excited(FE)_eigens}%
        (a) Energy spectrum of the system (colored diamonds and black dots \change{merging into horizontal and vertical} lines). The colors represent noticeable contributions of the eigenstates to the state $|\Psi_{f}\rangle$ ($w_m \geq 10^{-12}$) for the driving field with $\omega_p = 19.1$ and $\Phi_0 = 0.2$ [orange line in (b)]. (b) Evolution of the order parameter $\langle\hat{\eta}^2\rangle/L$ driven by the pulses~\eqref{eq:field} with the respective envelopes \change{of $\Phi(t)$} indicated by the dotted lines.
        } 
\end{figure}

Let $|\Psi_f\rangle$ denote the final state after the evolution driven by the pulse~\eqref{eq:field} with $\omega_p = 19.1$ \change{(this resonant frequency is slightly higher than in Ref.~\cite{Kaneko2020PRL} due to a difference in pulse envelopes)} and $\Phi_0 = 0.2$ [orange lines in Fig.~\ref{fig:excited(FE)_eigens}(b)]. This state can be expanded in the basis of eigenstates~\eqref{EqEigenstatesEigenenergies}, 
\begin{align}\label{eq:SpectralDecPsi}
    |\Psi_f\rangle = \sum_m \langle\psi_m|\Psi_{f}\rangle \ket{\psi_m}.
\end{align}
The diamonds in Fig.~\ref{fig:excited(FE)_eigens}(a) are color coded to visualize $w_m = |\langle\psi_m|\Psi_{f}\rangle|^2$, the probability of occupying eigenstate $\ket{\psi_m}$ after interacting with the laser pulse. The diamonds label only states with $w_m \geq 10^{-12}$. The colored diamonds are centered on the black dots that show the eigenenergies and the value of the order parameter. Note that the eigenstate 
with $\langle\hat{\eta}^2\rangle_m/L = 1.5$ and $\varepsilon_m \approx 56.28$ has an occupation probability greater than 50\%. However, we observe no population of the state with the highest-possible order parameter, $\langle\hat{\eta}^2\rangle_{\max}/L = (L/2+1)/2$ [see the black dot in the upper right corner of Fig.~\ref{fig:excited(FE)_eigens}(a)], which corresponds to the unique fully antisymmetric eigenstate (with respect to permutations between doublons and holons) with energy $\varepsilon_m = LU/2 = 80$. This suggests that utilizing quantum control could enhance the superconductivity. 

Dotted lines in Fig.~\ref{fig:excited(FE)_eigens}(b) show three envelopes of $\Phi(t)$ with $N_p = 54$, idle time $t_l = 5$ before the pulse [$t=\tau+t_l$ in Eq.~\eqref{eq:field}], and different values of $\omega_p$ and $\Phi_0$ along with the induced evolutions of the order parameter $\langle\hat{\eta}^2\rangle=\bra{\Psi(t)} \hat{\eta}^2 \ket{\Psi(t)}$ depicted by the solid lines.
\changeb{We also introduce the final time~$t_f$ at which we stop the measurement (or apply the quantum control), $t_f = t_l + T_p N_p + t_r$, with the same duration of idle time $t_r=5$ after the pulse.}
The depicted dependencies of the order parameter clearly show that pulses with certain resonance frequencies enhance superconducting properties, similar to results of Refs.~\cite{Kaneko2019PRL,Kaneko2020PRL}. The persistence of superconducting correlations after the laser field is turned off depends on the amplitude $\Phi_0$. The system becomes almost transparent for pulses with a relatively small \change{change in the carrier frequency}, as seen from the blue curve in Fig.~\ref{fig:excited(FE)_eigens}(b), which shows the dynamics for $\omega_p = 18.0$ and $\Phi_0 = 0.2$. 

\begin{figure}[tb]
    \includegraphics[width= \linewidth]{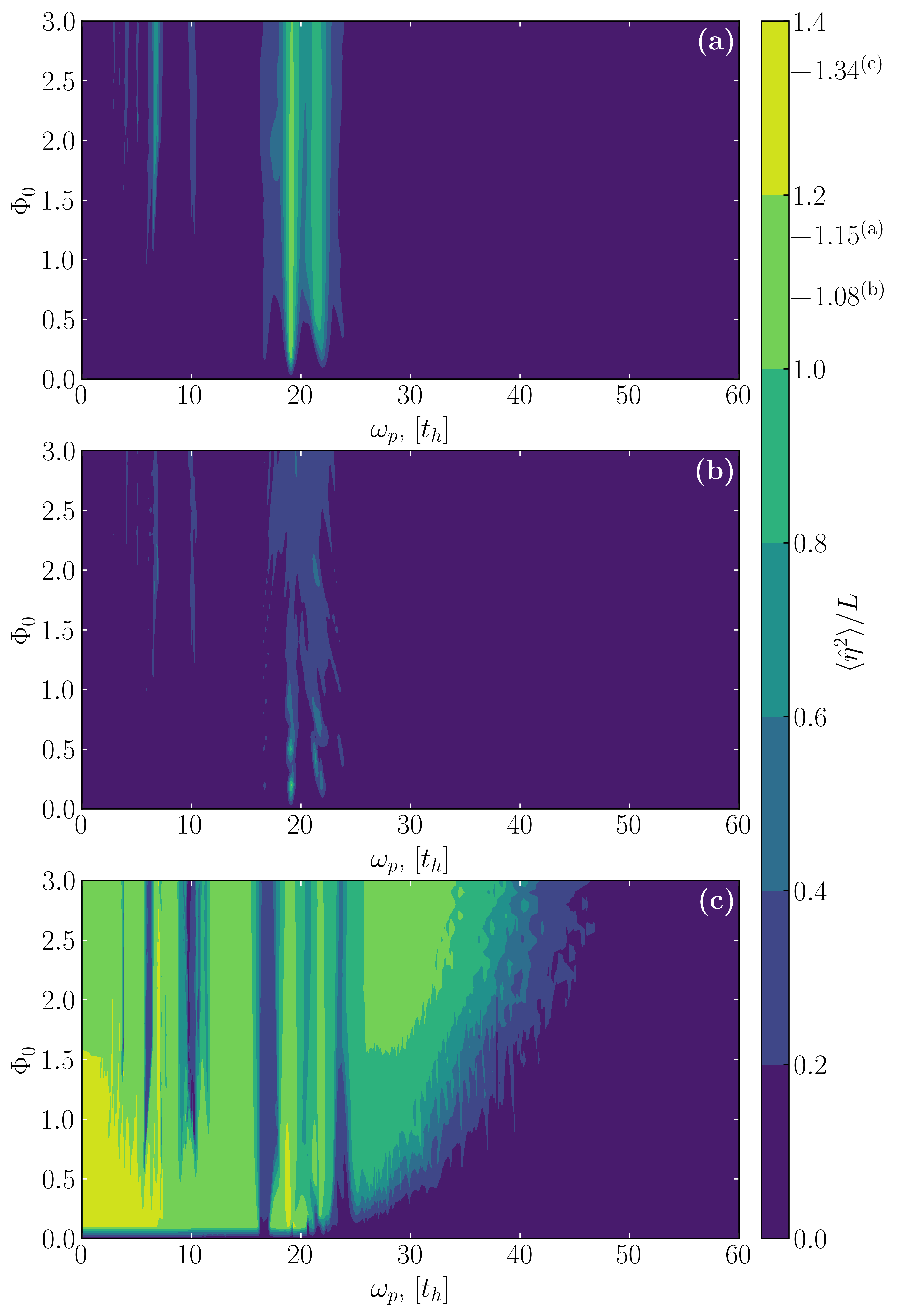}
    \caption{\label{fig:FE_2_grids}%
        (a) The maximal values of the order parameter achieved during evolution as functions of the pulse parameters $\omega_p$ and $\Phi_0$ (no quantum control). 
        (b) The \changeb{steady-state} values of the order parameter at the end of the same simulation \changeb{($t=t_f$,} no quantum control). (c) The \changeb{measured} values of the order parameter after the quantum Lyapunov control is  applied at  $t=t_f$ (see Sec.~\ref{subsec:LQC} for details).
        The maximal values achieved in each panel are indicated by the respective additional ticks in the color bar.
        }
\end{figure}

To analyze in detail the dependence of photoexcited superconducting correlations on a pulse's frequency and amplitude, we run a number of independent simulations for different values of $\omega_p$ or $\Phi_0$. The results of this parameter scan are summarized in \change{Figs.~\ref{fig:FE_2_grids}(a) and \ref{fig:FE_2_grids}(b)}.  The maximum value of the order parameter attained is shown in \change{Fig.~\ref{fig:FE_2_grids}(a)}.  The steady-state value of the order parameter, reached after the laser pulse is turned off, is depicted in \change{Fig.~\ref{fig:FE_2_grids}(b)}. A wide split-resonance band near $\omega_p \approx U = 20$ and several narrow side stripes at lower frequencies can be observed, whereas no significant excitations are seen in the high-frequency regime $\omega_p \geq 30$. \change{Qualitatively, we attribute the origin of the resonance splitting to finite-size effects and the one-dimensional character of the problem. In this system, the noninteracting density of states of mobile carriers (doublon and holon excitations) exhibits a minimum at the center and two maxima at the edges of the corresponding energy band.}

\section{Control of Superconducting correlations}\label{sec:QC}

\subsection{Enhancing  superconducting correlations with the \change{quantum Lyapunov} control}\label{subsec:LQC}
The \change{quantum Lyapunov} control~\cite{kosloff_excitation_1992, sugawara_control_1994, sugawara_control_1995, ohtsuki_application_1998, tannor_laser_1999, sugawara_general_2003, mirrahimi_reference_2005}, a quantum adaptation of the celebrated Lyapunov control  \cite{Lyapunov1992}, is a feedback protocol for updating the control field to maintain a non-negative time derivative of the target property. The Ehrenfest theorem for $\hat{\eta}^2$ reads
\begin{equation}\label{eq:Ehrenfest}
    \frac{\dd}{\dd t} \langle \hat{\eta}^2 \rangle = i \left\langle [\hat{\mathcal{H}},\hat{\eta}^2 ] \right\rangle.
\end{equation}
One obtains from Eqs.~\eqref{eq:hamiltonian} and \eqref{eq:eta_sq},
\begin{equation}\label{eq:commutator}
    [\hat{\mathcal{H}},\hat{\eta}^2 ] = - i t_h\sin[\Phi(t)]  \hat{\mathcal{Q}},
\end{equation}
where 
\begin{equation}\label{eq:Q}
    \hat{\mathcal{Q}} = \sum_{i,j} (-1)^{i+j} 
    \left\{ \hat{c}_{i,\uparrow}^{\dagger}
    \hat{c}_{i+1,\downarrow}^{\dagger} -\hat{c}_{i,\downarrow}^{\dagger} \hat{c}_{i+1,\uparrow}^{\dagger} 
    , \hat{c}_{j,\uparrow}^{} \hat{c}_{j,\downarrow}^{} 
    \right\} + \text{H.c.};
\end{equation}
hence,
\begin{equation}\label{eq:derivative}
    \frac{\dd}{\dd t}  \langle \hat{\eta}^2 \rangle = t_h \sin[\Phi(t)]  \langle   \hat{\mathcal{Q}} \rangle.
\end{equation}
The key idea behind the Lyapunov control is that with a judicious choice of $\Phi(t)$, we can ensure $\frac{\dd}{\dd t} \langle \hat{\eta}^2 \rangle \geq 0$, thereby guaranteeing that the superconducting order parameter $\langle \hat{\eta}^2 \rangle$ does not decrease over time. We choose the control field to be
\begin{equation}\label{eq:Phi_LE}
    \Phi^{\rm{(L)}}\Big\{|\Psi(t)\rangle\Big\} = \arcsin \frac{\langle\hat{\mathcal{Q}}\rangle}{{\mathcal{Q}}_{\max}} 
    \Longrightarrow \frac{\dd}{\dd t} \langle \hat{\eta}^2 \rangle \geq 0,
\end{equation}
where \change{${\mathcal{Q}}_{\max}$ is the largest absolute value of the eigenvalue of $\hat{\mathcal{Q}}$ to ensure} that the argument of $\arcsine$ is in $[-1, +1]$.


Note that it is not sufficient to just replace $\Phi(t)$ in the Hamiltonian~\eqref{eq:hamiltonian} with the control pulse $\Phi^{\rm{(L)}}\{|\Psi(t)\rangle\}$ and launch the quantum evolution, since this leads to the trivial result $\Phi^{\rm{(L)}}\{|\Psi(t)\rangle\} = 0$, i.e., no laser field, and $\langle \hat{\eta}^2 \rangle = 0$. Therefore, it is necessary first to excite the system with, say, \change{an initial} pulse~\eqref{eq:field} followed by the Lyapunov control pulse~\eqref{eq:Phi_LE}. We will denote such a concatenated pulse by 
\begin{align}\label{eq:ConcatenatedPulse}
    \Phi_{\rm{L}}(t) 
    = 
    \begin{cases}
        \Phi(t) \text{ from Eq.~\eqref{eq:field}}, & \text{if } t < t^{\rm{(act.)}}, \\
        \Phi^{\rm{(L)}}\{|\Psi(t)\rangle\}, & \text{if } t \geq t^{\rm{(act.)}}.
    \end{cases}
\end{align}
We have empirically found that the Lyapunov protocol~\eqref{eq:Phi_LE} should replace the monochromatic driving~\eqref{eq:field} when the time-derivative of the order parameter averaged over the preceding period of monochromatic field oscillations $T_p = 2\pi/\omega_p$ turns negative, i.e.,
\begin{equation}\label{eq:LE_condition}
    \left \langle \frac{\dd}{\dd t}  \langle \hat{\eta}^2 \rangle \right \rangle_{T_p} 
    = \frac{t_h}{T_p} \int_{t^{\rm{(act.)}} - T_p}^{t^{\rm{(act.)}}}  \sin[\Phi(t)]  \langle   \hat{\mathcal{Q}} \rangle dt
    < 0.
\end{equation} 
This condition is robust to numerical noise that can manifest in tiny negative values of $\frac{\dd}{\dd t}  \langle \hat{\eta}^2 \rangle$. Other threshold-based conditions to \changeb{activate} the Lyapunov control can also be employed; e.g., we activate $\Phi^{\rm{(L)}}\{|\Psi(t)\rangle\}$ when $\frac{\dd}{\dd t}  \langle \hat{\eta}^2 \rangle < -10^{-3}$. 

To demonstrate the superiority of the Lyapunov control over simple monochromatic driving, Fig.~\ref{fig:FE_2_grids}(c) shows the values of the superconducting order parameter reached at the end of evolution driven by the concatenated field~\eqref{eq:ConcatenatedPulse} as a function of the pump pulse's~\eqref{eq:field} frequency $\omega_p$ and amplitude $\Phi_0$ before it is replaced by the Lyapunov control pulse once the activation condition~\eqref{eq:LE_condition} is satisfied. Note that the final value of the order parameter is also the maximum value attained since the Lyapunov control ensures that the order parameter does not decrease [see Eq.~\eqref{eq:Phi_LE}]. Comparing Figs.~\ref{fig:FE_2_grids}(b) and \ref{fig:FE_2_grids}(c) reveals that the Lyapunov control becomes especially efficient at low frequencies and small amplitudes of the pump pulse.

\begin{figure}[tb]
    \includegraphics[width=\linewidth]{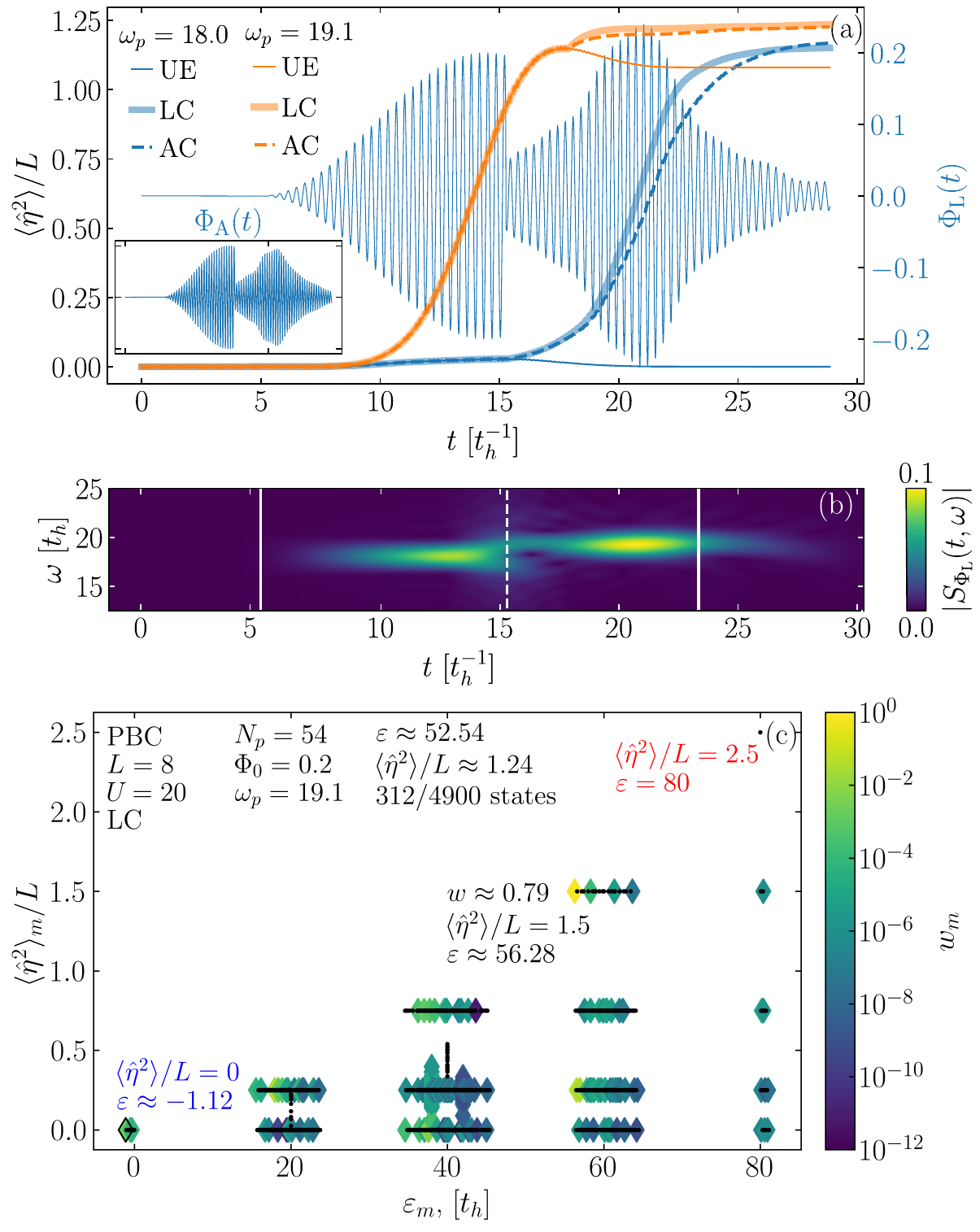}
    \caption{\label{fig:LQC_AQC}%
        (a) Evolution of $\langle\hat{\eta}^2\rangle/L$ under $\Phi(t)$ \change{(solid thin lines)}, $\Phi_{\rm{L}}(t)$ \change{[solid thick lines (LC)]}, and $\Phi_{\rm{A}}(t)$ [dashed lines (AC); see Sec.~\ref{asymptotic} for details]. The driving fields~$\Phi_{\rm{L}}(t)$ (main panel) and $\Phi_{\rm{A}}(t)$ (inset) are shown only for $\omega_p = 18.0$. \change{(b) Magnitude of the short-time Fourier transform $|S_{\Phi_{\mathrm{L}}}(\omega,t)|$, illustrating the temporal evolution of the spectrum $\Phi_{\mathrm{L}}(t)$. Regions outside the solid vertical lines correspond to border effects, where the analysis windows extend beyond the pulse duration. The dashed line indicates $t^{(\mathrm{act.})}$. (c)} The contributions of 312 eigenstates~\eqref{EqEigenstatesEigenenergies} (color-coded diamond symbols) to the final state with energy $\varepsilon\approx52.54$ and the order parameter $\langle\hat{\eta}^2\rangle/L\approx1.24$ achieved during the LC evolution for the initial field $\Phi(t)$ with $\omega_p = 19.1$ and $\Phi_0 = 0.2$.
        }
\end{figure}

To gain further insights into the evolution driven by the concatenated field~\eqref{eq:ConcatenatedPulse} with the activation condition~\eqref{eq:LE_condition}, we separately analyze in Fig.~\ref{fig:LQC_AQC}(a) the time evolution of $\langle\hat{\eta}^2\rangle$ for two monochromatic pulses from Fig.~\ref{fig:excited(FE)_eigens}(b) that drive the weakest ($\omega_p = 18.0$, $\Phi_0 = 0.2$) and strongest ($\omega_p = 19.1$, $\Phi_0 = 0.2$) excitations of the order parameter. 
The \changeb{thin solid lines} in Fig.~\ref{fig:LQC_AQC}(a) depict the evolutions of $\langle\hat{\eta}^2\rangle$ driven by monochromatic drivings [uncontrolled evolution (UE)]. The \changeb{thick (slightly lighter) solid} lines depict the evolutions of $\langle\hat{\eta}^2\rangle$ achieved via the \change{quantum Lyapunov control (LC)}. Not only does the Lyapunov control significantly enhance the weakest off-resonant steady-state value (blue), but it also leads to a slight increase of the previously obtained maximal value in the resonant case (orange).

To understand how the Lyapunov control leads to the enhancement of the superconducting correlations in the resonant case, 
\change{Fig.~\ref{fig:LQC_AQC}(c)} [similar to Fig.~\ref{fig:excited(FE)_eigens}(a)] depicts the population of  eigenstates~\eqref{EqEigenstatesEigenenergies} at the end of the dynamics driven by the concatenated pulse~\eqref{eq:ConcatenatedPulse} with $\omega_p = 19.1$ and $\Phi_0 = 0.2$ for the initial monochromatic pump pulse $\Phi(t)$. As in Fig.~\ref{fig:excited(FE)_eigens}(a), $w_m = |\langle\psi_m|\Psi_{f}\rangle|^2$ is obtained from Eq.~\eqref{eq:SpectralDecPsi}, where $\ket{\Psi_f}$ is the final state. Comparing the outcome of the pure monochromatic driving [Fig.~\ref{fig:excited(FE)_eigens}(a)] with the Lyapunov control [\change{Fig.~\ref{fig:LQC_AQC}(c)}], we observe that for the Lyapunov control twice as many eigenstates have $w_m \geq 10^{-12}$; additionally, the eigenstate with $\langle\hat{\eta}^2\rangle/L = 1.5$ and $\varepsilon_m\approx56.28$ has increased its population to more than 75\%. Surprisingly, no contribution is observed from the eigenstate with the largest-possible order parameter [see the black dot in the upper right corner of \change{Fig.~\ref{fig:LQC_AQC}(c)} as well as Fig.~\ref{fig:excited(FE)_eigens}(a)]. 

\change{
Figure~\ref{fig:LQC_AQC}(b) shows that the Lyapunov control automatically adjusts the carrier frequency from the initial off-resonant value (e.g., $\omega_{p}=18.0$) to the optimal frequency corresponding to the lowest-frequency resonance ($\omega_{r,1}\approx19.1$), thereby enhancing superconducting correlations. 
We verified that this behavior occurs across wide ranges of initial frequencies~$\omega_p$ and amplitudes~$\Phi_0$. The observed modulations of the pulse envelope and carrier frequency are experimentally feasible based on current implementations (see, e.g., Ref.~\cite{rey-de-castro_time-resolved_2013}).
}
\changeb{Note also that the rapid change in the pulse envelope at the moment of activation of the Lyapunov control, as in Fig.~\ref{fig:LQC_AQC}(a), can be smoothed with additional filters with a minimal impact on the overall enhancement of superconducting correlations (see Appendix~\ref{app:A} for details).}

\subsection{When is it best to activate the Lyapunov control?}\label{subsec:locality}

\begin{figure}[tb]
    \includegraphics[width= \linewidth]{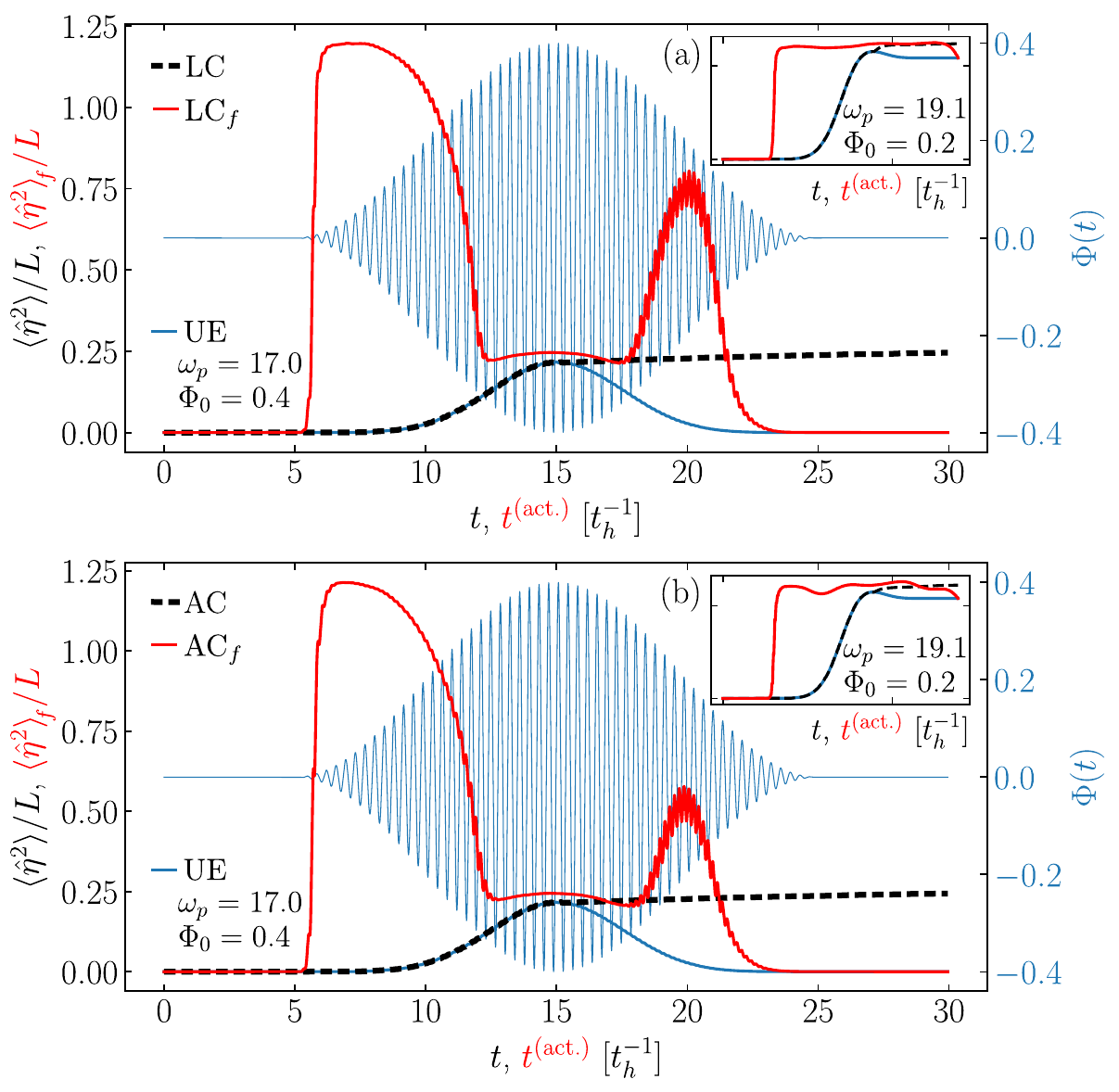}
    \caption{\label{fig:nonlocality}%
        {(a) The \changeb{final} value of $\langle\hat{\eta}^2\rangle/L$ \changeb{at $t=t_f$} with respect to activation time $t^{\rm{(act.)}}$ [red curves (\changeb{$\textrm{LC}_f$})] along with the evolution of $\langle\hat{\eta}^2\rangle/L$ driven by $\Phi(t)$ [blue \changeb{lines (UE)}] and $\Phi_{\rm{L}}(t)$ with the switching condition~\eqref{eq:LE_condition} [black dashed lines (LC)]. The uncontrolled driving field is shown only in the main panel ($\omega_p=17.0$, $\Phi_0 = 0.4$); the results in the inset correspond to $\omega_p=19.1$ and $\Phi_0 = 0.2$. (b) The same quantities as in (a), but red solid (\changeb{$\textrm{AC}_f$}) and black dashed lines correspond to \change{quantum asymptotic control (AC}; see Sec.~\ref{asymptotic}).}
        }
\end{figure}

The evolution induced by the concatenated field~\eqref{eq:ConcatenatedPulse} depends on the activation time $t^{\rm{(act.)}}$—when the monochromatic field is replaced by the Lyapunov control. Up to this point, the activation time $t^{\rm{(act.)}}$ was chosen by the condition~\eqref{eq:LE_condition}. However, Fig.~\ref{fig:FE_2_grids}(c) contains many examples in which such a switching condition does not enhance the order parameter (see, in particular, the case of the pump pulse with $\omega_p=17.0$ and $\Phi_0 = 0.4$). Nevertheless, changing $t^{\rm{(act.)}}$ improves the \changeb{final} value of the order parameter \changeb{at time $t=t_f$}. The red curve (\changeb{$\textrm{LC}_f$}) in the main panel of Fig.~\ref{fig:nonlocality}(a) depicts \changeb{this} value of $\langle\hat{\eta}^2\rangle$ as a function of $t^{\rm{(act.)}}$ for the pump pulse with $\omega_p=17.0$ and $\Phi_0 = 0.4$.

Recall that from Figs.~\ref{fig:excited(FE)_eigens} and \ref{fig:LQC_AQC}, the monochromatic excitation with $\omega_p=19.1$ and $\Phi_0 = 0.2$ yielded a high value of the order parameter. In such a case, the value of the order parameter $\langle\hat{\eta}^2\rangle$ reached at the end of the evolution driven by the concatenated pulse~\eqref{eq:ConcatenatedPulse} as a function of the activation time $t^{\rm{(act.)}}$, shown in the inset in Fig.~\ref{fig:nonlocality}(a), reveals considerable freedom in choosing $t^{\rm{(act.)}}$.
Note that in both cases analyzed in Fig.~\ref{fig:nonlocality}(a), it is sufficient to turn on the Lyapunov control just after a few oscillations of the initial pump field to achieve a significant amplification of the superconducting order parameter. This seems natural since activating the control earlier provides more time for the \changeb{enhancement of superconducting correlations before their measurement at the fixed time~$t_f$; see also Appendix~\ref{app:A} for additional details.}

\subsection{Asymptotic quantum control}\label{asymptotic}

The maximum possible value of the superconducting order parameter is $\langle\hat{\eta}^2\rangle_{\max}/L = (L/2 + 1)/2 =  2.5$, yet the most we have got so far is  $\langle\hat{\eta}^2\rangle/L \approx 1.34$ [see Fig.~\ref{fig:FE_2_grids}(c)].  It is tempting to modify the Lyapunov control protocol to see whether $\langle\hat{\eta}^2\rangle_{\max}$ can be reached.

Let us introduce the \change{quantum \emph{asymptotic control} (AC)}. Unlike the Lyapunov control~\eqref{eq:derivative} that aims to increase $\langle\hat{\eta}^2\rangle$, AC tends to decrease the difference between $\langle\hat{\eta}^2\rangle$ and some chosen target value $\eta_{0}^2$, which can be set to $\langle\hat{\eta}^2\rangle_{\max}$. Since
\begin{equation}\label{eq:derivative_A}
    \frac{\dd}{\dd t} 
    \left(
            \langle\hat{\eta}^2\rangle - \eta^{2}_0
    \right)^2   
    = 2 \left(\langle\hat{\eta}^2\rangle - \eta^{2}_0\right) t_h \sin[ \Phi(t) ]  \langle   \hat{\mathcal{Q}} \rangle,
\end{equation}
to ensure that this derivative is non-positive, it is sufficient to choose a control $\Phi(t)$ in the form
\begin{align}\label{eq:Phi_AE}
    \Phi^{\rm{(A)}}\Big\{|\Psi(t)\rangle \Big\} &= - \arcsin 
    \frac{
        \langle\hat{\mathcal{Q}}\rangle
        \left(\langle\hat{\eta}^2\rangle - \eta^{2}_0\right)
    }
    {{\mathcal{Q}}_{\max} \eta^2_{\max}} \\
    &\Longrightarrow \frac{\dd}{\dd t} 
    \left(
            \langle\hat{\eta}^2\rangle - \eta^{2}_0
    \right)^2 \leq 0, \notag
\end{align}
where \change{${\mathcal{Q}}_{\max}$ and $\eta^2_{\max}$ are the largest absolute values of the eigenvalues of $\hat{\mathcal{Q}}$ and $\hat{\eta}^2$, respectively, to ensure} that the argument of $\arcsine$ is in the range $[-1, +1]$.

As in the case of LC, we start the evolution from the ground state for which $\langle\hat{\mathcal{Q}}\rangle=0$; hence an initial monochromatic pump pulse is needed to kick the dynamics. Also, we need to choose the activation time $t^{\rm{(act.)}}$ when the pump should be replaced by the asymptotic control~\eqref{eq:Phi_AE}. The activation condition~\eqref{eq:LE_condition},  employed in Sec.~\ref{subsec:LQC}, leads to very similar results in the case of AC, as shown in Fig.~\ref{fig:LQC_AQC}(a). As shown in the inset of Fig.~\ref{fig:LQC_AQC}(a), the resulting concatenated field
\begin{align}
    \Phi_{\rm{A}}(t) =
    \begin{cases}
        \Phi(t) \text{ from Eq.~\eqref{eq:field}}, & \text{if } t < t^{\rm{(act.)}}, \\
        \Phi^{\rm{(A)}}\{|\Psi(t)\rangle\}, & \text{if } t \geq t^{\rm{(act.)}},
    \end{cases}
\end{align}
is very similar to $\Phi_{\rm{L}}(t)$ [Eq.~\eqref{eq:ConcatenatedPulse}]. Also, the dependence of the final value of $\langle\hat{\eta}^2\rangle$ on the parameters of the pump pulse is almost identical to the one attained by the Lyapunov control shown in Fig.~\ref{fig:FE_2_grids}(c). 

Finally, AC leads to a marginal improvement of the final maximum value of the order parameter $\langle\hat{\eta}^2\rangle/L\approx1.37$ over the Lyapunov control result of $\langle\hat{\eta}^2\rangle/L\approx1.34$, still significantly falling short of $\langle\hat{\eta}^2\rangle_{\max}$. Perhaps to further increase the superconducting correlations, one needs to resort to a much more computationally expensive approach of optimal quantum control, which, unlike both the Lyapunov and asymptotic controls, typically requires a large number of iterations to converge the unknown control. 

We should note that the excitability of superconducting correlations depends more strongly on the switching condition determining $t^{\rm{(act.)}}$ (see Sec.~\ref{subsec:locality}) than on the specific type of the employed quantum control, i.e., LC or AC. Similar to the \change{quantum Lyapunov} control, asymptotic control does not make condition~\eqref{eq:LE_condition} effective for the monochromatic pump pulse with $\omega_p=17.0$ and $\Phi_0 = 0.4$. Figure~\ref{fig:nonlocality}(b) (main panel) shows the dependence of the \changeb{final} value of $\langle\hat{\eta}^2\rangle/L$ on $t^{\rm{(act.)}}$ (red solid line) in the case of AC [compare with LC in Fig.~\ref{fig:nonlocality}(a)].

\subsection{Suppressing   superconducting correlations with the \change{quantum Lyapunov} control}\label{sec:Suppressing}

\begin{figure}[tb]
    \includegraphics[width= \linewidth]{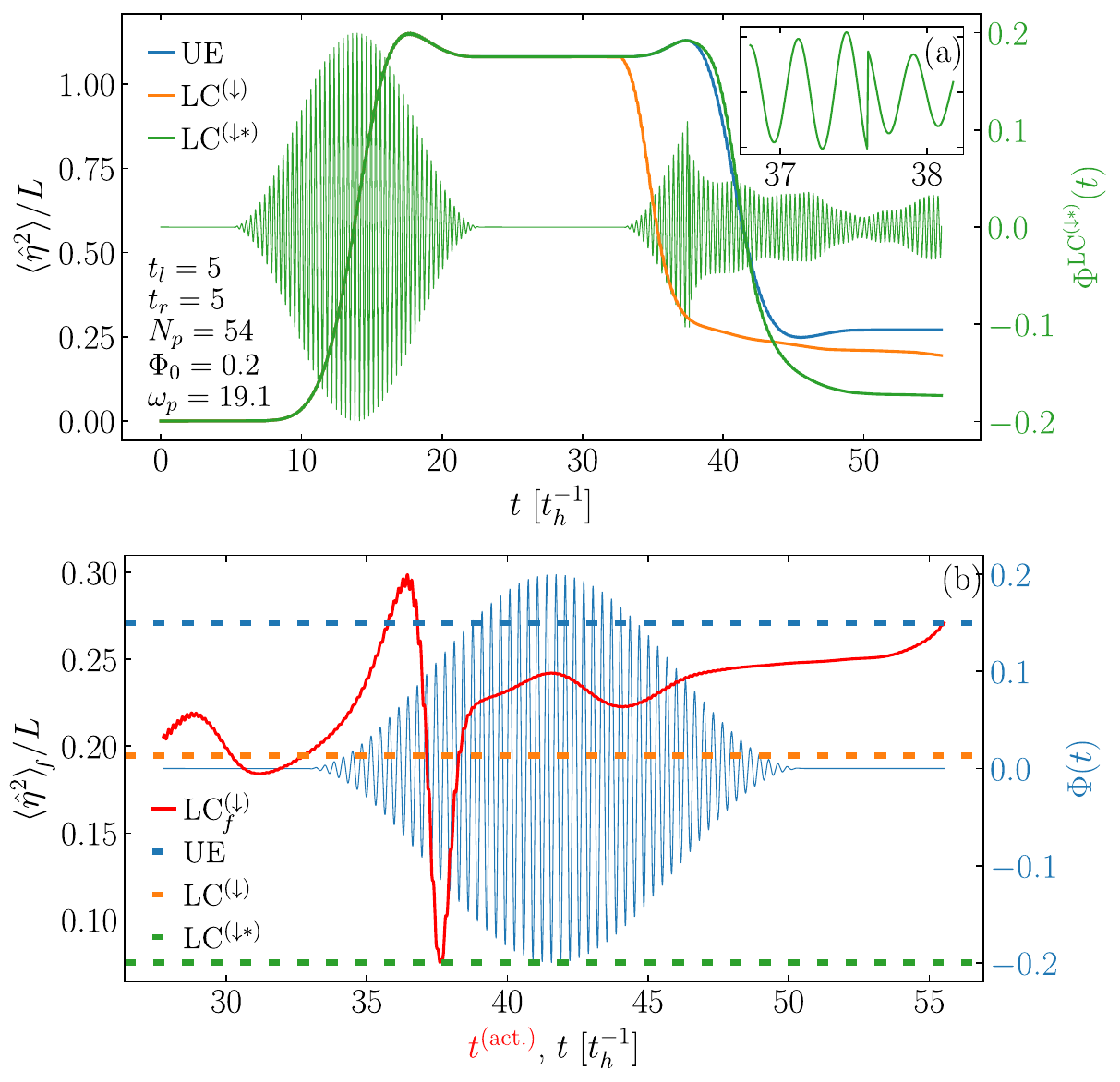}
    \caption{\label{fig:suppression}%
        (a) {Excitation and consequent suppression of the superconducting correlations with $\Phi^{\rm (UE)}(t)$ from Eq.~\eqref{eq:PhiUE} (labeled UE), $\Phi^{\rm{LC^{(\downarrow)}}}(t)$ from Eq.~\eqref{eq:LQCdown} ($\rm{LC^{(\downarrow)}}$), and $\Phi^{\rm{LC^{(\downarrow *)}}}(t)$ with the optimal $t^{\rm{(act.)}}$ ($\rm{LC^{(\downarrow *)}}$).} 
        Parameters are $\omega_p = 19.1$ and $\Phi_0 = 0.2$. 
        (b) The value of $\langle\hat{\eta}^2\rangle/L$ \changeb{measured at $t=t_f$} as a function of $t^{\rm{(act.)}}$ (red line) and the  \changeb{corresponding} values reached in UE, $\rm{LC^{(\downarrow)}}$, and $\rm{LC^{(\downarrow *)}}$ evolutions (dashed lines). 
        }
\end{figure}
In Sec.~\ref{subsec:LQC}, we established that the \change{quantum Lyapunov} control enhances superconducting correlations. We now demonstrate that the opposite can also be achieved—the suppression of superconducting correlations, thereby enabling almost full on-demand control of quantum correlations.
 
From Eq.~\eqref{eq:derivative}, it readily follows that the following control accomplishes the suppression of superconductivity:
\begin{equation}\label{eq:Phi_LE_sup}
    \Phi^{\downarrow}\Big\{|\Psi(t)\rangle\Big\} = - \arcsin \frac{\langle\hat{\mathcal{Q}}\rangle}{{\mathcal{Q}}_{\max}}
    \Longrightarrow \frac{\dd}{\dd t}\langle\hat{\eta}^2\rangle\leq 0.
\end{equation}
Note that the controls for enhancing [Eq.~\eqref{eq:Phi_LE}] and suppressing [Eq.~\eqref{eq:Phi_LE_sup}] the correlations differ only by sign. 

To benchmark the Lyapunov control for suppressing superconductivity, we first illustrate the effect of monochromatic driving. We take the pump pulse $\Phi(t)$ from Eq.~\eqref{eq:field} with parameters $\omega_p = 19.1$ and $\Phi_0 = 0.2$. This pulse yields $\langle\hat{\eta}^2\rangle/L \approx 1.08$ -- the largest \changeb{steady-state value of} the order parameter attained in Fig.~\ref{fig:excited(FE)_eigens}(b). The blue line (labeled UE) in Fig.~\ref{fig:suppression}(a) shows the evolution of the order parameter when the pulse $\Phi(t)$ is applied twice resulting in the field
\begin{align}\label{eq:PhiUE}
    \Phi^{\rm (UE)}(t) = \Phi(t) + \Phi(t - {\Delta t}),
\end{align}
The first pulse induces the superconductive correlations. When the second identical pulse is applied at a later time ($\Delta t=t_l+t_r+T_pN_p$, where $t_l$ and $t_r$ are the idle times before and after the pulse, respectively), the order parameter decreases to $\langle\hat{\eta}^2\rangle/L \approx 0.25$ and ultimately settles at $\langle\hat{\eta}^2\rangle/L \approx 0.27$.

As illustrated by the orange line (labeled $\rm{LC^{(\downarrow)}}$) in Fig.~\ref{fig:suppression}(a), the following Lyapunov control allows us to further decrease  $\langle\hat{\eta}^2\rangle/L \approx 0.19$:
\begin{align}\label{eq:LQCdown}
    \Phi^{\rm{LC^{(\downarrow)}}}(t) =
    \begin{cases}
        \Phi^{\rm (UE)}(t) \text{ from Eq.~\eqref{eq:PhiUE}}, & \text{if } t < t^{\rm{(act.)}}, \\
        \Phi^{\downarrow}\Big\{|\Psi(t)\rangle\Big\}, & \text{if } t \geq t^{\rm{(act.)}},
    \end{cases}
\end{align}
where the activation time $t^{\rm{(act.)}}$ of the Lyapunov suppressing control~\eqref{eq:Phi_LE_sup} is chosen such that
\begin{align}
    t^{\rm{(act.)}} \geq {\Delta t}, \qquad
     \int_{t^{\rm{(act.)}} - T_p}^{t^{\rm{(act.)}}}  \sin[\Phi(t)]  \langle   \hat{\mathcal{Q}} \rangle dt > 0. \label{eq:LE_anti-condition}
\end{align}
Note that the latter condition is directly inspired by  Eq.~\eqref{eq:LE_condition}.

However, we can push the order parameter even further down to $\langle\hat{\eta}^2\rangle/L \approx 0.08$. The final value of $\langle\hat{\eta}^2\rangle$ achieved strongly depends on the choice of the activation time $t^{\rm{(act.)}}$. This dependency is visualized by the red curve in Fig.~\ref{fig:suppression}(b). The minimum of the red curve gives the optimal choice of the activation time. The control field~\eqref{eq:LQCdown} with such a choice of $t^{\rm{(act.)}}$ yields $\langle\hat{\eta}^2\rangle/L \approx 0.08$. The full time evolution of the order parameter is shown as the green curve (labeled $\rm{LC^{(\downarrow *)}}$) in Fig.~\ref{fig:suppression}(a). Note that the order parameter first reaches a maximum before the descent. This strong suppression comes at the cost of an abrupt change in the control field at the activation time, as shown in the inset of Fig.~\ref{fig:suppression}(a). \changeb{The abrupt change in the control field can be smoothed with additional filters, as we discuss in Appendix~\ref{app:A}.}

\change{\section{Conclusion and Outlook}}

We demonstrated that \change{quantum Lyapunov} control provides an effective method for enhancing superconducting correlations in the one-dimensional Fermi-Hubbard model without requiring intricate fine tuning of pulse parameters. Our key findings include the following: 
First, traditional monochromatic driving excites superconducting correlations only within narrow frequency ranges, making it challenging to optimize without extensive parameter searches. In contrast, Lyapunov control efficiently amplifies even weakly excited states, particularly at low frequencies and amplitudes of the initial pumping field.

Second, from the theoretical perspective, our approach eliminates the need for precise knowledge of the optimal carrier frequency and the pulse envelope since they are automatically adjusted during time evolution from the system response. This may also be used as a tool for finding the first (lowest-frequency) resonance in the system.
Third, we showed that a variant of this approach—asymptotic quantum control—yields a comparable enhancement of superconducting correlations, confirming the robustness of time-local control methods for this application. 
Finally, we demonstrated that Lyapunov control can also be used to effectively suppress previously induced superconducting correlations, enabling bidirectional control of quantum correlations on demand.
These results highlight the potential of \change{quantum Lyapunov} control as a practical tool for manipulating superconducting properties in strongly correlated electron systems.

\change{
The derived quantum Lyapunov control~\eqref{eq:Ehrenfest}--\eqref{eq:Phi_LE} remains valid for nonzero temperature states if the averages $\langle\hat{\eta}^2\rangle = \operatorname{Tr}(\hat{\rho}\hat{\eta}^2)$ and $\langle\hat{\cal Q}\rangle = \operatorname{Tr}(\hat{\rho}\hat{\cal Q})$ are defined with respect to the density matrix $\hat{\rho}(t)$, whose unitary evolution is governed by the von Neumann equation. However, as discussed in Ref.~\cite{Kaneko2020PRL}, one must analyze both the correlation function $\langle\hat{\eta}^2\rangle$ and the charge stiffness since the condition $\langle\hat{\eta}^2\rangle > 0$ does not guarantee superconducting current in nonzero temperature.

The developed Lyapunov control can be further extended to include dissipation by using the Lindblad master equation. In this case, system dynamics and quantum control efficiency will depend strongly on dissipation. These directions offer promising opportunities for both theoretical studies and experimental realizations.
}

\vspace{1cm}

\change{
\section*{Code and data availability}
The data that support the findings of this article are openly available~\footnote{\url{https://github.com/OleksandrPovitchan/SC-pairing-Q-control}}.
}

\acknowledgments

D.I.B is grateful for the support from the Army Research Office (ARO) (grant W911NF-23-1-0288; program manager Dr.~James Joseph). A.G.S. acknowledges the financial support in the framework of the IMPRESS-U grant from the US National Academy of Sciences via STCU project No.~7120 and the program for young researchers from the National Academy of Sciences of Ukraine No.~139/2024-07. A.G.S. is also grateful for the support from the Carol Lavin Bernick Faculty Grant Program at Tulane University.

The views and conclusions contained in this document are those of the authors and should not be interpreted as representing the official policies, either expressed or implied, of ARO or the U.S. Government. The U.S. Government is authorized to reproduce and distribute reprints for Government purposes notwithstanding any copyright notation herein. 

\appendix
\change{
\section{Smooth activation \changeb{and deactivation} of quantum control}\label{app:A}

\begin{figure}[tb]
    \includegraphics[width= \linewidth]{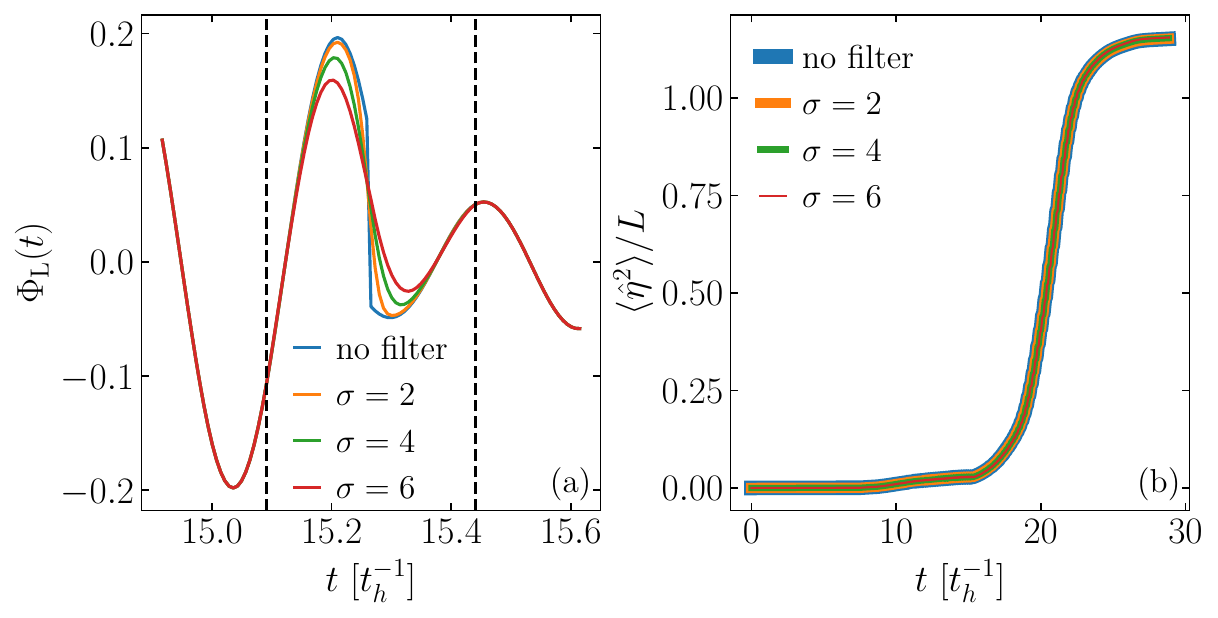}
    \caption{\label{fig:smooth}%
        \change{(a) {The effect of applying a Gaussian filter with three different standard deviations $\sigma$ to the original discontinuous pulse used in Fig.~\ref{fig:LQC_AQC}(a). The view is zoomed in on the LC activation region; the area between vertical dashed lines indicates where the filter was applied.} 
        (b) {The nearly identical responses demonstrate the robustness of the quantum Lyapunov control to smoothing of discontinuities.}}
        }
\end{figure}

In Figs.~\ref{fig:LQC_AQC}(a) and \ref{fig:suppression}(a), one can observe abrupt changes in field amplitudes once quantum control is activated. We demonstrate that smoother, more experimentally feasible amplitude changes do not alter the main results regarding control and enhancement of pairing correlations.

We use the same parameters and quantum Lyapunov control procedure shown in Fig.~\ref{fig:LQC_AQC}(a) for the field with $\omega_p=18.0$. In the time interval $t\in[t^{(\text{act})}-T_p/2,t^{(\text{act})}+T_p/2]$, we apply a modified Gaussian filter.
Figure~\ref{fig:smooth} shows that changing to a smoother activation protocol does not significantly affect the quantum control procedure's efficiency.
}


\changeb{Another relevant question arises when analyzing the behavior of the control field near the final measurement time, $t_f$.
In particular, Figs.~\ref{fig:LQC_AQC}(a) and \ref{fig:suppression}(a) show that the quantum control field does not fully vanish within the chosen time window, which corresponds to the duration of one and two consecutive uncontrolled pulses, respectively, with idle times $t_l = t_r = 5$.

From a physical standpoint, in the dissipationless model~\eqref{eq:hamiltonian} the order parameter should remain unchanged once the field is completely switched off, since at $\Phi = 0$ we have $[\hat{\mathcal{H}}(0), \hat{\eta}^2] = 0$, in accordance with Eq.~\eqref{eq:commutator}. However, its precise value may still depend on the details of how the pulse is switched off.

\begin{figure}
    \includegraphics[width=\linewidth]{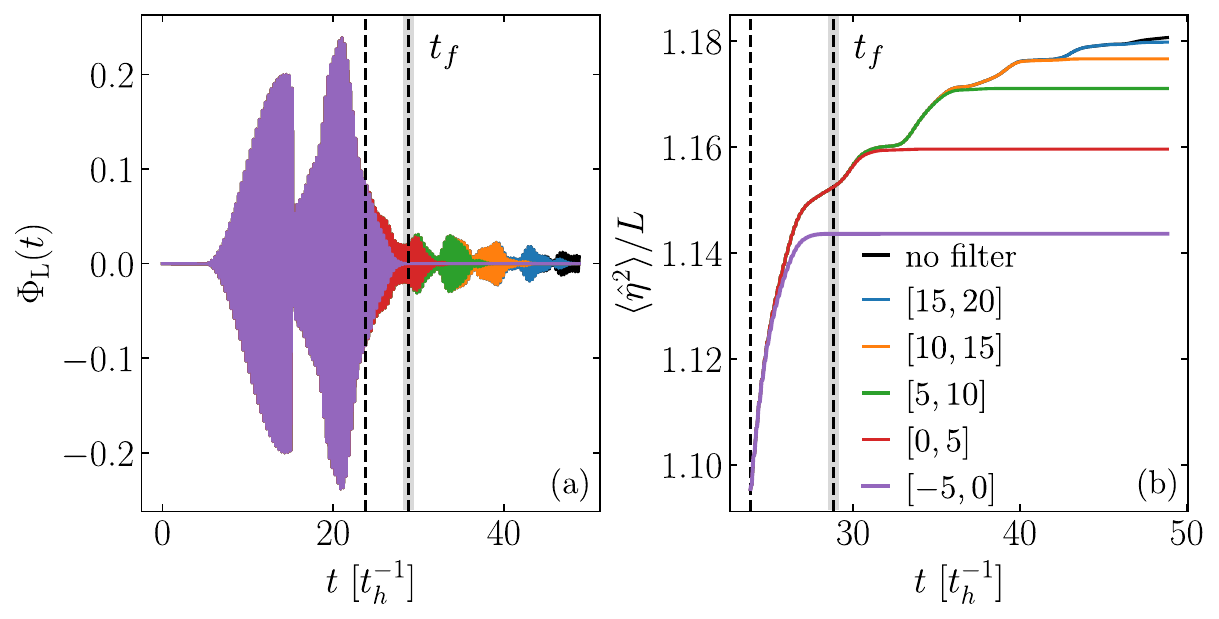}
    \caption{\label{fig:switch_off}%
        \changeb{(a) Control field $\Phi_{\mathrm{L}}(t)$ from Fig.~\ref{fig:LQC_AQC}(a), extended beyond $t_f$, for the initial pulse with $\omega_p=18.0$ and $\Phi_0=0.2$. (b) Dependence of the order parameter on the choice of the deactivation interval. 
        In both panels, the colored curves are labeled by the time intervals $[\tau_1, \tau_2]$ that specify the parameters $t_{1,2} = \tau_{1,2} + t_f$  of the switch-off function~\eqref{eq:filter}. The vertical dashed lines show the interval $[-5, 0]$.}
        }
\end{figure}

In the extended time frame in Fig.~\ref{fig:switch_off} [compare Fig.~\ref{fig:LQC_AQC}(a) with $\omega_p=18.0$], we observe that after a substantial drop in the amplitude of $\Phi_{\mathrm{L}}(t)$, the order parameter continues to increase at a much reduced rate, gradually saturating for $t>50$.
The switch-off can be accelerated by smoothly truncating the tail of the controlled field over a chosen interval $[t_1,t_2]$ as $\Phi_{\mathrm{L}}(t) \to \Phi_{\mathrm{L}}(t) \Xi(t)$ with the switch off function 
\begin{align}\label{eq:filter}
    \Xi(t) = 
    \begin{cases}
        1, & \text{if } t < t_1, \\
        \cos^2\!\left[\dfrac{\pi}{2}\,\dfrac{t - t_1}{t_2 - t_1}\right], & \text{if } t_1 \le t < t_2,\\ 
        0, & \text{if } t \ge t_2.
    \end{cases}
\end{align}

Figure~\ref{fig:switch_off}(b) shows that the resulting values of the order parameter $\langle \hat{\eta}^2\rangle$ after a smooth deactivation of $\Phi_{\mathrm{L}}(t)$ are robust, exhibiting only a weak dependence on the chosen deactivation interval within the extended time frame. This confirms the overall stability of our results with respect to reasonable choices of the turn-off function.}

\bibliography{references}
\end{document}